\documentclass [a4paper, 11pt, twocolumn, twoside]{article}

\usepackage{amsmath, amsfonts, amssymb, latexsym}
\usepackage{cancel}

\usepackage{color}
\usepackage[dvipsnames]{xcolor}
\usepackage{graphics, graphicx}
\usepackage{tikz} % use specific libraries, if needed
\usepackage{subcaption} % subfigures
\usepackage{array, booktabs, makecell} % tables
\usepackage{dblfloatfix} 

\usepackage{fancyhdr}
\usepackage{multicol, multirow}

\usepackage{listings}

\usepackage[hidelinks]{hyperref}
\usepackage{breakurl}
\usepackage{authblk}

\usepackage{siunitx}

\usepackage[textsize=tiny, textwidth=2.4cm]{todonotes}

\usepackage{ngerman}    % (doesn't hurt even for English texts)
\usepackage{ulem} % durchstreichen

\newcommand{\ConfName}{\textit{Scientific Computing 2023}}

\title{Finding the Optimum Design of Large Gas Engines Prechambers Using CFD and Bayesian Optimization}
\author[1]{Stefan Posch}
\author[1]{Clemens Gößnitzer}
\author[2]{Franz Rohrhofer}
\author[2]{Bernhard C. Geiger}
\author[1,3]{Andreas Wimmer}
\affil[1]{\small LEC GmbH}
\affil[2]{\small Know-Center GmbH}
\affil[3]{\small Institut für Thermodynamik und nachhaltige Antriebssysteme, TU Graz}
\date{} % to be omitted for the proceedings

\newcommand{\authorshort}{Posch et al.}
\newcommand{\titleshort}{Prechamber Bayesian Design Optimization}

\begin{document}

\maketitle\thispagestyle{plain}

\begin{abstract}
The turbulent jet ignition concept using prechambers is a promising solution to achieve stable combustion at lean conditions in large gas engines, leading to high efficiency at low emission levels. Due to the wide range of design and operating parameters for large gas engine prechambers, the preferred method for evaluating different designs is computational fluid dynamics (CFD), as testing in test bed measurement campaigns is time-consuming and expensive. However, the significant computational time required for detailed CFD simulations due to the complexity of solving the underlying physics also limits its applicability. In optimization settings similar to the present case, i.e., where the evaluation of the objective function(s) is computationally costly, Bayesian optimization has largely replaced classical design-of-experiment. Thus, the present study deals with the computationally efficient Bayesian optimization of large gas engine prechambers design using CFD simulation. Reynolds-averaged-Navier-Stokes simulations are used to determine the target values as a function of the selected prechamber design parameters. The results indicate that the chosen strategy is effective to find a prechamber design that achieves the desired target values.
\end{abstract}	

\section{Introduction} 
\label{sec:intro}
Internal combustion engines (ICE) will continue to play an important role in the generation of energy and the transportation of goods. The high energy density of the used fuels in combination with the robust and reliable technology, the ICE will still be the first choice for applications in which pure electrified systems make little sense as it is the case for the shipping of goods. Furthermore, ICE will be a key part for future energy management by the reconversion of chemical energy storage media such as hydrogen into electricity or to ensure grid stability which is induced by the increasing use of alternative energy sources such as wind and solar energy~\cite{Guelpa2019}. Hence, the research on increasing engine efficiencies under the fulfilment of stringent emission legislation will continue regardless of whether classical carbon-based or alternative carbon-free fuels are used. In the case of spark ignited (SI) engines, a common approach is the operation under lean conditions which increases the engine efficiency by increasing the specific heat capacity of the working fluid, reducing heat losses through the cylinder walls, and reducing pumping losses in part load operation~\cite{Zhu2022}. However, the operation at lean conditions comes with the drawback of unstable combustion and thus, high cyclic fluctuations that can even lead to misfiring cycles. The enhancement of the ignition energy by increasing the performance of the ignition system helps to overcome these issues. A well-established concept to increase the ignition efficiency of lean mixture SI engines is the turbulent jet ignition (TJI) by the use of a separate, small (compared to the main combustion chamber) volume known as prechamber. The mixture is ignited by the spark plug in the prechamber and the flame propagates through the volume increasing the pressure and force hot products and active radicals to flow into the main combustion chamber via overflow bores. Thus, the ignition of the mixture in the main combustion chamber is induced at multiple locations due to the penetrating turbulent jets coming from the prechamber. However, the design of a prechamber and the choice of the appropriate operating conditions to achieve a highly efficient TJI system in combination with the main combustion chamber is not straightforward due to the large number of parameters and thus degrees of freedom. Therefore, experimental investigations as shown by Roethlisberger and Favrat~\cite{Roethlisberger2003} or Novella et al.~\cite{Novella2020} are complex as well as cost intense and thus limit the number of design variants that can be tested. To decrease the experimental effort, the state-of-the-art procedure is to apply numerical simulation, i.e. computational fluid dynamics (CFD), to increase the number of design and operation parameter variants by virtual testing and to achieve a comprehensive view on the physical effects which measurements are not able to provide. 

\begin{figure}
\centering
\includegraphics[height=7cm, trim={2cm 5cm 2cm 0.5cm},clip]{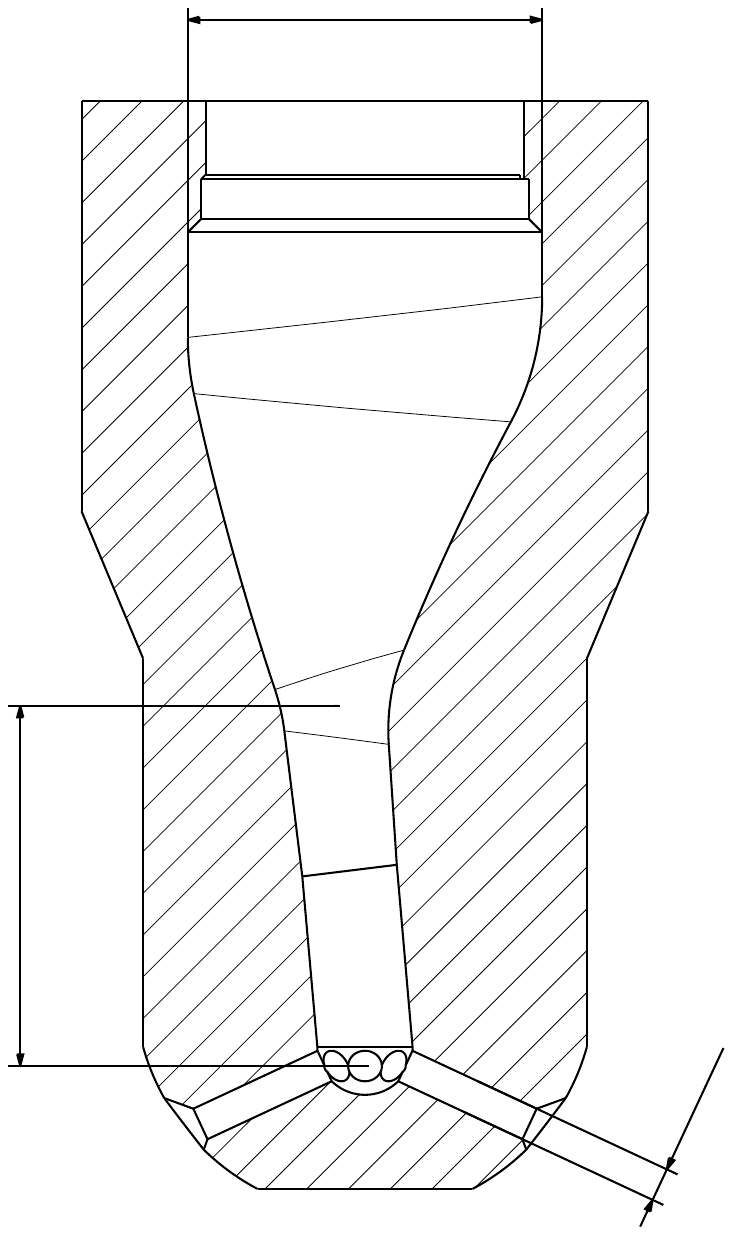}
\begin{picture}(0,0)(0,0)
\put(-89,185){{$d_{bottle}$}}
\put(-140,45){\rotatebox{90}{$h_{neck}$}}
\put(-45,28){\rotatebox{65}{$d_{bore}$}}
\end{picture}
\caption{Base prechamber design and the three input parameters}\label{fig:design_parameters}
\end{figure}

Various studies regarding the application of CFD to evaluate prechamber efficiencies and related values can be found in literature. Benajes et al.~\cite{Benajes2020} investigated the scavenging and turbulence distributions of several prechamber designs under ultra-lean conditions. The results of the study indicate that the reduction of the laminar flame speed due to the ultra-lean conditions significantly decreases the quality of the turbulent jets exiting the prechamber. Silva et al.~\cite{Silva2020} carried out CFD combustion simulations to investigate the influence of several design parameters on the engine combustion characteristics. The authors stated that the diameter of the overflow bores impacts the peak pressure and the residence time of the main charge in the prechamber. Furthermore, the results show a large influence of the neck diameter on the pressure increase in the prechamber as well es the aforementioned residual time. Related works can be found in Feng et al.~\cite{Feng2018}, Winter et al. ~\cite{Winter2018}, or Zhang et al.~\cite{Zhang2020}.
In addition to common CFD simulation, data-driven approaches and machine learning (ML) concepts are increasingly finding their way into the design optimization of ICE components to facilitate the discovery of appropriate prechamber configurations. Ge et al.~\cite{Ge2021} compared different ML models to develop a surrogate model which represents combustion CFD simulations. The trained model is further used in a genetic algorithm to find an optimum prechamber design. To decrease the number of required CFD simulations, a Bayesian updating optimization method was applied. In a similar study, Silva et al.~\cite{Silva2022} compared design of experiments (DOE) optimization with different ML surrogate models and combinations of optimization techniques via a so-called automated super-learner framework. The results indicate that the proposed method performance is superior compared to the classical DOE approach. Although these studies use ML surrogate models for the optimization process, the training of the models requires a certain number of combustion CFD simulations which are computationally expensive. To overcome this issue, Posch et al.~\cite{Posch2021} presented a method to use condensed CFD simulations to train ML models, which are able to predict trends in the prechamber performance, thus helping to reduce the computational effort.
Following the aforementioned studies of Ge et al.~\cite{Ge2021} and Silva et al.~\cite{Silva2022}, the present study deals with the optimization of a prechamber design by the use of CFD and appropriate optimization techniques. A Bayesian optimization (BO) algorithm based on Gaussian processes is applied which is initialized by a DOE-based CFD data set. To reduce the computational effort, the CFD simulation only covers the range from bottom dead center until ignition timing (IT) rather than the entire cycle. This simplification can be justified as follows: (i) The present study focuses solely on the prechamber behaviour and thus gas exchange through the valves in the prechamber is neglected. (ii) Since the flow field around the spark plug at IT has a significant influence on the flame kernel development~\cite{GhaderiMasouleh2019, Goesnitzer2021, Posch2022} and furthermore on the combustion in the pre- and main chamber, the optimization process focuses on the turbulent kinetic energy and velocity magnitude around the spark plug at IT. The results show that the proposed method is able to find a prechamber design that covers three design parameters and that fulfills the target values in terms of maximum turbulent kinetic energy by considering a certain maximum velocity magnitude around the spark plug. Figure~\ref{fig:design_parameters} shows the base prechamber design with the three input parameters for the optimization process.

%%%%%%%%%%%%%%%%%%%%%%%%%%%%%%%%%%%%%%%%%%%%%%%%%%%%%%%%%%%%%
\section{Methodology} 
\label{sec:methodology}

\subsection{Engine specification}
\label{ssec:Engine_specification}

The engine considered here was adapted from an existing engine. The main chamber was simplified to a perfect cylinder, and the prechamber had variable geometrical sizes as described later.  The simplified engine had a bore of \SI{190}{mm}, a stroke of \SI{220}{mm}, a displaced volume of about \SI{6.2}{L} and a speed of \SI{1500}{rpm}.  No valves or gas exchange were modelled.  Thus, the time of interest for this study covered the engine cycle ranging from bottom dead centre at \SI{540}{CAD} until ignition timing at \SI{695}{CAD}.

\subsection{CFD setup}
\label{ssec:CFD_setup}

The Reynolds-averaged Navier-Stokes (RANS) CFD simulations were performed with OpenFOAM (version from November 2022~\cite{OpenFOAMCommit} from The OpenFOAM Foundation~\cite{OpenFOAM}).  Meshing was done with snappyHexMesh.  The grid size was \SI{4}{mm} in the cylinder, \SI{0.25}{mm} in the prechamber and \SI{0.125}{mm} in the prechamber bores and near the spark plug, leading to approximately 1 mio cells.  Turbulence was modelled by the RANS $k$-$\omega$ SST two-equation model.  Initial temperature and pressure were set to \SI{400}{K} and \SI{3}{bar}, respectively.  The time step was adapted to keep the Courant number below \num{2}.
The PISO algorithm with three correctors was used for pressure-velocity coupling.  Additionally, two non-orthogonal correctors were used to improve convergence of the pressure prediction.  Furthermore, second-order bounded discretization schemes were used for gradient and divergence, whereas a first-order implicit time discretization was used.  Since the maximum observed speeds exceed a Mach number of \num{0.3}, the transonic option was set for all simulations. Mesh movement was done with the standard OpenFOAM layered engine approach.
Mass conservation was monitored by integration of the density over the volume, to check if the total mass of the closed volume remains constant, which indeed was the case for all simulations. A more detailed description of the simulation setup can be found in~\cite{Gossnitzer2022}. For the evaluation of the  quantities near the spark plug, all cells within a sphere with radius of \SI{0.5}{mm} just outside of the spark plug gap were monitored.  The relevant simulation output, namely the turbulence intensity measured by the volume-average value of $k$ and velocity magnitude $|\vec{v}|$\, are both evaluated inside this sphere.

\subsection{Optimization strategy}
\label{ssec:Optimization_strategy}

Three geometrical parameters of the prechamber were adjusted in this study to find the optimal shape: the diameter of the prechamber bottle ($d_{\mathrm{bottle}}$), the diameter of the overflow bores ($d_{\mathrm{bore}}$), and the height of the prechamber neck ($h_{\mathrm{neck}}$).
The optimization ranges were chosen as $d_{\mathrm{bottle}}\in[8,12]$, $d_{\mathrm{bore}}\in[0.75, 1.15]$ and $h_{\mathrm{neck}}\in[15,20]$. 

From the perspective of optimization, and summarizing the geometrical parameters of the prechamber in the tuple $\mathbf{x}=(d_{\mathrm{bottle}},d_{\mathrm{bore}},h_{\mathrm{neck}})$, the aim is to find a set of solutions to
\begin{equation}\label{eq:optimization}
    \max_\mathbf{x} k(\mathbf{x}) \quad \text{s.t.} \quad |\vec{v}(\mathbf{x})|\le 25.
\end{equation}
In BO~\cite{Frazier2018}, both the objective $k$ and the constraint $|\vec{v}|$ are considered as black box functions that are expensive to evaluate, and that are thus approximated by surrogates. The most common surrogate models are Gaussian processes~\cite{RasmussenWilliams}. Given a dataset  $\mathcal{D}=\{(\mathbf{x}_i,k(\mathbf{x}_i),|\vec{v}(\mathbf{x}_i)|)\}$ of different simulated settings, the Gaussian process models yield (Gaussian) distributions $\hat k$ and $|\hat{\vec{v}}|$ for the target $k$ and the constraint $|\vec{v}|$ for every possible parameterization $\mathbf{x}$. If $\mathbf{x}$ is close to a point $\mathbf{x}_i$ in the dataset $\mathcal{D}$, then the distribution of, e.g., $\hat k(\mathbf{x})$ has a mean close to $k(\mathbf{x}_i)$ and a small variance, whereas if $\mathbf{x}$ is far from every point in $\mathcal{D}$, the variance of $\hat k(\mathbf{x})$ will be large --- there is high \emph{epistemic uncertainty}. During optimization, BO relies on these probabilistic models $\hat k(\mathbf{x})$ and $|\hat{\vec{v}}(\mathbf{x})|$ to determine (using acquisition functions, see~\cite{Frazier2018}) candidate geometries $\mathbf{x}^\bullet$ that improve upon all geometries in $\mathcal{D}$ in the sense of~\eqref{eq:optimization}. The thus proposed candidate geometries are then evaluated using CFD simulations, and consequently yield a new tuple $(\mathbf{x}^\bullet,k(\mathbf{x}^\bullet),|\vec{v}(\mathbf{x}^\bullet)|)$. This tuple is included in $\mathcal{D}$, and the probabilistic models $\hat k(\mathbf{x})$ and $|\hat{\vec{v}}(\mathbf{x})|$ are updated. This optimization loop is iterated until either a predetermined number of iterations is exceeded or until the utility of the solution is satisfactory.

To address the optimization problem~\eqref{eq:optimization}, we select $\mathbf{x}^\bullet$ via the expected constrained improvement~\cite{Gardner2014}. 
% BG: The following paragraph can be removed if space is needed
Specifically, if $\mathbf{x}_j$ is the geometry in $\mathcal{D}$ that maximizes $\hat k(\mathbf{x})=k(\mathbf{x})$, then $\mathbf{x}^\bullet$ is chosen as an optimizer of
\begin{multline}\label{eq:qCEI}
        \max_\mathbf{x} \mathbb{E}\left(\mathbf{1}(|\hat{\vec{v}}(\mathbf{x})|\le 25)\cdot \max\{0,\hat k(\mathbf{x})-k(\mathbf{x}_j)\}\right)\\
        =\max_\mathbf{x} PF(\mathbf{x})\cdot EI(\mathbf{x})
\end{multline}
where first factor is the probability that the candidate $\mathbf{x}$ is feasible (i.e., that $|\hat{\vec{v}}(\mathbf{x})|\le 25$) and where the second factor is the standard expected improvement (w.r.t.\ $k$).
% paragraph end
Furthermore, to exploit the fact that CFD simulations can be run in parallel, batch (or parallel) BO was used, which responds with a set of $q$ candidate solutions $\{\mathbf{x}^\bullet_1,\dots,\mathbf{x}^\bullet_q\}$, cf.~\cite[eq.~(14)]{Frazier2018}

In this work, BO was implemented using BoTorch~\cite{BoTorch}, a Python library that contains functionality for constrained, multi-objective, and batch BO. Specifically, both $\hat k$ and $|\hat{\vec{v}}|$ are modeled as Gaussian processes with a fixed noise standard deviation of \num{0.005} and a Matern kernel with optimized hyperparameters. 
For optimal functionality of the BO, the three  geometry parameters $\mathbf{x}$ are scaled to the closed unit interval $[0, 1]$ and the objectives are scaled to unit variance ($k$ and $|\vec{v}|$) and zero mean (only $k$).
The expected constrained improvement was evaluated using Monte Carlo sampling, with \num[]{1024} samples. The batch size was set to five.

A DOE study was performed with ten samples to create the initial data for three consecutive BO iterations based on latin hypercube formulation. The optimum value found by the DOE study was $k=160.38$ (see Table~\ref{tab:results}).

%%%%%%%%%%%%%%%%%%%%%%%%%%%%%%%%%%%%%%%%%%%%%%%%%%%%%%%%%%%%%
%------- Figure Begin ------%
\begin{figure*}[!b]
\centering
\begin{subfigure}{\textwidth}
    \includegraphics[width=\textwidth]{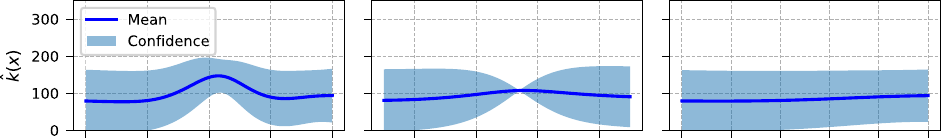}
    \caption{DOE Initialization (10 data points)}
    \label{fig:doe_init}
\end{subfigure}
\par\medskip
\begin{subfigure}{\textwidth}
    \includegraphics[width=\textwidth]{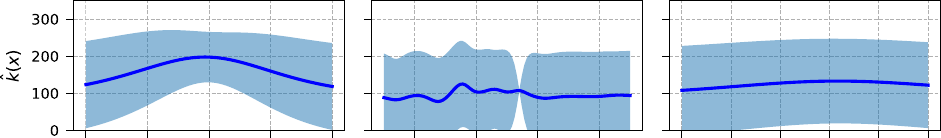}
    \caption{First BO Iteration (15 data points)}
    \label{fig:first_BO}
\end{subfigure}
\par\medskip
\begin{subfigure}{\textwidth}
    \includegraphics[width=\textwidth]{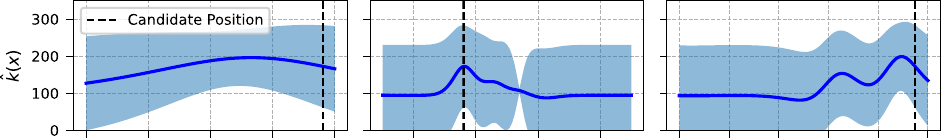}
    \caption{Second BO Iteration (20 data points)}
    \label{fig:second_BO}
\end{subfigure}
\par\medskip
\begin{subfigure}{\textwidth}
    \includegraphics[width=\textwidth]{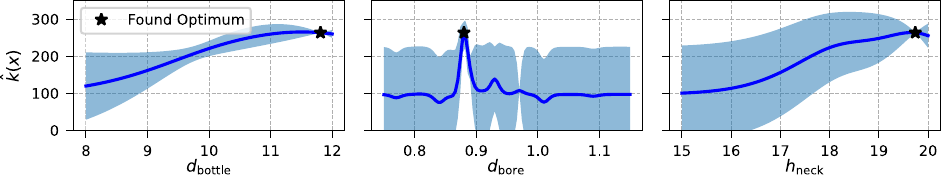}
    \caption{Third BO Iteration (25 data points)}
    \label{fig:third_BO}
\end{subfigure}       
\caption{Progress of three consecutive BO iterations. The three columns show the mean function of $\hat k(\mathbf{x})$ and its standard deviation for two of the dimensions of $\mathbf{x}$ fixed to $\mathbf{x}^\star$, respectively}
\label{fig:training_optimum}
\end{figure*}
%-------- Figure End -------%
\section{Results}
\label{sec:results}
% Results table
The results for the optimum values found by the DOE initialization and three consecutive BO iterations can be found in Table~\ref{tab:results}.
While the first BO iteration shows considerable improvement ($k=246.46$) compared to the DOE initialization, the second BO iteration was not able to further increase the objective within the following five candidate positions ($k=244.56$).
The third and final BO iteration again yields a new optimum value ($k=263.16$) while still fulfilling the constraint ($|\vec{v}|=22.53$). 

%Description of Figure 2
Figure~\ref{fig:training_optimum} illustrates the progress of BO over the three iterations. Given that after the third iteration the optimal value in $\mathcal{D}$ is $\mathbf{x}^\star=(d_{\mathrm{bottle}}^\star,d_{\mathrm{bore}}^\star,h_{\mathrm{neck}}^\star)$, the three columns show the mean function of $\hat k(\mathbf{x})$ and its standard deviation for two of the dimensions of $\mathbf{x}$ fixed to $\mathbf{x}^\star$, respectively. One can see that by including more training data, the surrogate $\hat k$ becomes more and more structured and that, especially w.r.t.\ $d_{\mathrm{bottle}}$, the epistemic uncertainty reduces.
The working of BO is best illustrated in the third
and fourth rows of the figure:
Figure~\ref{fig:second_BO} depicts the surrogate model $\hat k$ trained on 20 data points. The vertical dashed lines indicate the candidate position $\mathbf{x}^\bullet$ that maximizes the acquisition function~\eqref{eq:qCEI}. It can be seen that the candidate position lies at a local optimum of the mean of $\hat k$ for $d_{\mathrm{bore}}$, with a value close to 180. Running CFD simulations with the thus suggested geometry parameters $\mathbf{x}^\bullet$ yields a value of $k(\mathbf{x}^\bullet)$ exceeding 250. Figure~\ref{fig:third_BO} then shows the surrogate $\hat k$ after including $\mathbf{x}^\bullet$ (and the remaining four geometries of the batch) in the training data $\mathcal{D}$. It can be seen that training was successful in the sense that the $\mathbf{x}^\bullet=\mathbf{x}^\star$ now lies on the mean curve of $\hat k$ and that the epistemic uncertainty at this position is small.

%------- Table Begin ------%
\begin{table}[t!]
\caption{Optimum values found by the DOE initialization (10 data points) and three consecutive BO iterations (5 data points per iteration).}
\label{tab:results}
% for avoiding siunitx using bold extended
\renewrobustcmd{\bfseries}{\fontseries{b}\selectfont}
\renewrobustcmd{\boldmath}{}
% abbreviation
\newrobustcmd{\B}{\bfseries}
\resizebox{\columnwidth}{!}{
\begin{tabular}{c|cc|ccc}
\Xhline{3\arrayrulewidth}
& $k$ & $|\vec{v}|$ & $d_{\mathrm{bottle}}$ & $d_{\mathrm{bore}}$ & $h_{\mathrm{neck}}$ \\ \hline \hline
DoE Init. & 160.38 & 17.93 & 10.20 & 0.89 & 18.75 \\
$1^{\mathrm{st}}$ BO It. & 246.46 & 22.70 & 10.02 & 0.88 & 18.26 \\
$2^{\mathrm{nd}}$ BO It. & 244.56 & 22.11 & 9.90 & 0.88 & 18.18 \\
\B $3^{\mathrm{rd}}$ BO It. & \B 263.16 & \B 22.53 & \B 11.81 & \B 0.88 & \B 19.74 \\
\Xhline{3\arrayrulewidth}
\end{tabular}
}
\end{table}
%-------- Table End -------%

% Description of limitations; move/split to where you believe it fits best!
Despite the fact that BO is a global optimization method, we have observed in our experiments that most candidate geometries were suggested in the vicinity of optima of the surrogate function. In other words, BO \emph{exploited} knowledge about the currently best geometry rather than \emph{exploring} geometries that have not been simulated so far. Upon inspection of Figure~\ref{fig:training_optimum}, one can see that the target $k(\mathbf{x})$ varies substantially with the bore diameter; with only few training data, this variation cannot be captured in the surrogate $\hat k(\mathbf{x})$. (Indeed, the surrogate $\hat k$ in Figure~\ref{fig:second_BO} underestimated the value of the $k$ at the candidate position indicated with the dashed lines, as can be seen by the star in Figure~\ref{fig:third_BO}.) This is aggravated by the fact that the kernel parameters are optimized based on $\mathcal{D}$ and, thus, may not be optimal for the black box functions $k$ and $|\vec{v}|$. Indeed, as the center image in Figure~\ref{fig:second_BO} suggests, the automatic hyperparameter optimization may have terminated with a kernel width that is too large to capture the variation of $k$ with the bore diameter (especially when compared with the corresponding images in Figure~\ref{fig:first_BO} and~\ref{fig:third_BO}, respectively). A manual setting of these hyperparameters, however, is not possible since the functions $k$ and $|\vec{v}|$ are unknown a priori.

\begin{figure}
\centering
\includegraphics[width=8cm, trim={12cm 8cm 7cm 8cm},clip]{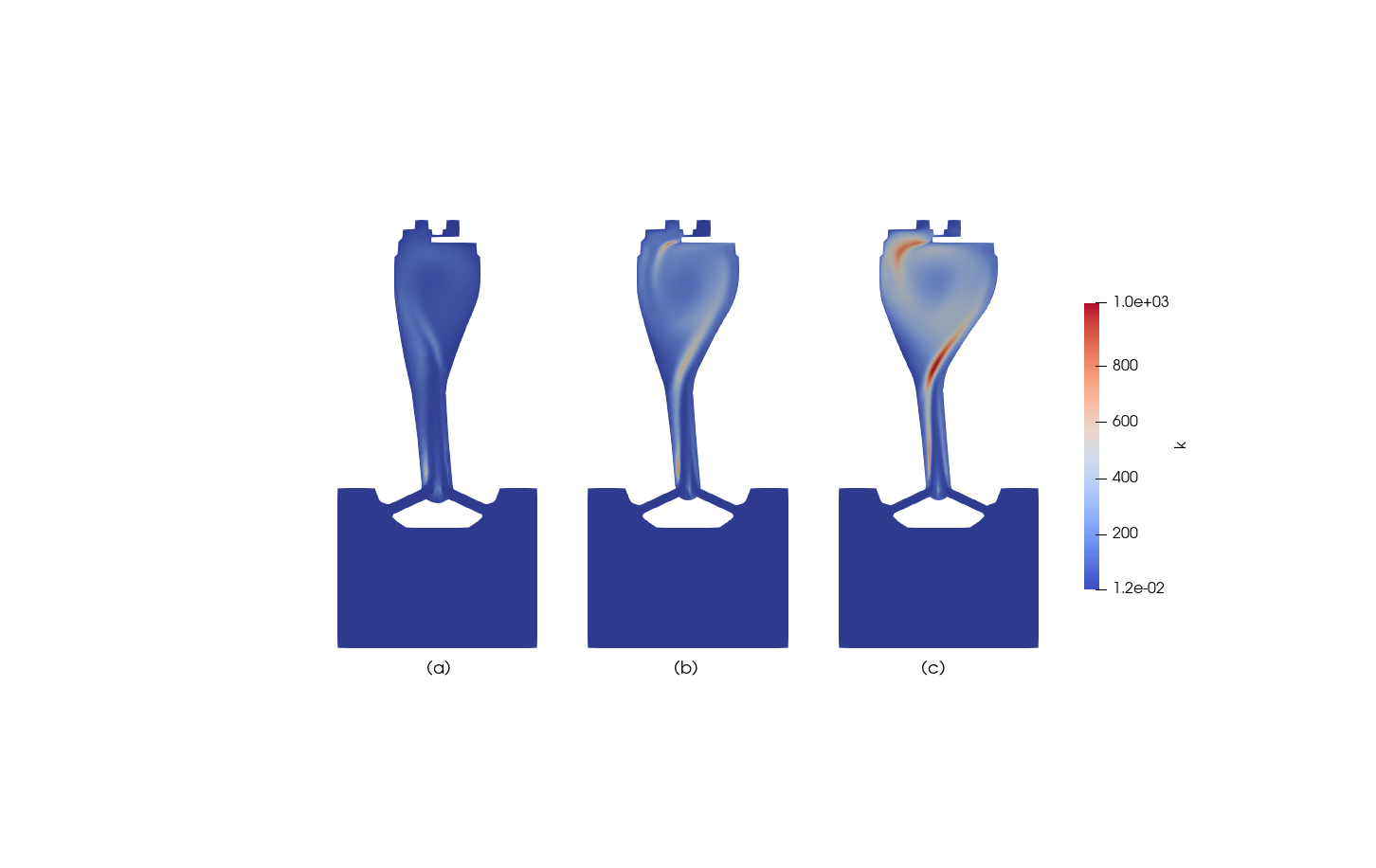}
\begin{picture}(0,0)(0,0)
\put(-90,0){{(a)}}
\put(-25,0){{(b)}}
\put(40,0){{(c)}}
\end{picture}
\caption{Turbulence intensity $k$-field in the prechamber cross section of one unfavourable design combination from the initial set (a), the best design combination of the initial set (b) and the final BO predicted design}\label{fig:k_CFD}
\label{fig:CFD_results}
\end{figure}
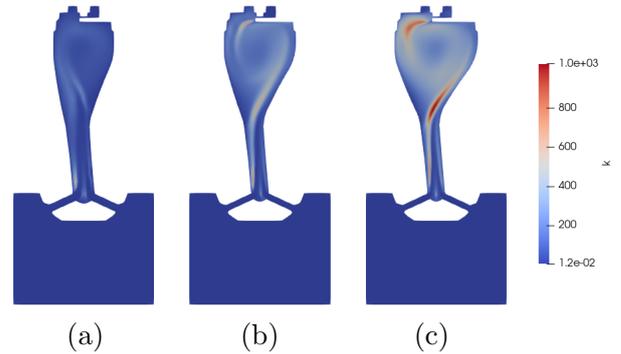

Figure~\ref{fig:CFD_results} shows the determined $k$-field in the cross section of the prechamber which allows a detailed analysis of the BO design parameter optimization results. While Figure~\ref{fig:CFD_results}a and Figure~\ref{fig:CFD_results}b represent an unfavourable and a favourable solution of the initial DOE set, respectively, Figure~\ref{fig:CFD_results}c shows the final resulting parameter combination found by the BO. It can be clearly observed, that the formation of a vortex in the prechamber is promoted by the appropriate combination of the chosen design parameters. Since the mass flowing from the main chamber into the prechamber is mainly influenced by the piston movement, a smaller value of $d_{\mathrm{bore}}$ results in higher overflow velocity and consequently higher turbulence velocities in the prechamber for the predefined parameter range. It should be mentioned at this point that this effect is not unlimited since the reduction of the overflow area can lead to choking effects. The combination a small $d_{\mathrm{bore}}$ value and high values of $d_{\mathrm{bottle}}$  as well as $h_{\mathrm{neck}}$ leads to a well-developed vortex with maximum values of $k$ in the vicinity of the spark plug. 

%------- Figure Begin ------%
%-------- Figure End -------%

%%%%%%%%%%%%%%%%%%%%%%%%%%%%%%%%%%%%%%%%%%%%%%%%%%%%%%%%%%%%%
\section{Summary and Outlook}
\label{sec:summary_outlook}

In the present study, a combination of CFD and BO was applied to optimize the design of a large bore gas engine prechamber. Three design parameter namely the diameter of the prechamber bottle, the diameter of the overflow bores and the height of the prechamber neck were chosen as input parameters. The target of the optimization process was maximizing the turbulence intensity $k$ under a velocity magnitude constraint in the vicinity of the spark plug. After the initialization of the system via ten DOE-based samples, three BO steps were carried out. The analysis of the BO indicate that both the strong variability of $k$ with the bore diameter and the effects of automatic hyperparameter optimization appear to slow down BO. One may therefore assume that substantially more iterations are required for BO to enter the exploration phase and, consequently, suggest geometry parameters distinct from the current optima. An option to speed up BO is to chose acquisition functions that permit influencing the trade-off between exploration and exploitation, such as the upper confidence bound. Future work shall be devoted to investigate these alternatives. Furthermore, additional investigations regarding the complexity of the physical system and its sensitivity to design parameter changes will be performed in order to increase the input and output parameter space.

\section*{Acknowledgments}

The authors acknowledge the financial support of the Austrian COMET — Competence Centers for Excellent Technologies — Programme of the Austrian Federal Ministry for Climate Action, Environment, Energy, Mobility, Innovation and Technology, the Austrian Federal Ministry for Digital and Economic Affairs, and the States of Styria, Upper Austria, Tyrol, and Vienna for the COMET Centers LEC EvoLET and Know-Center, respectively.
%The authors would like to acknowledge the financial support of the “COMET Module LEC HybTec” within the “COMET - Competence Centers for Excellent Technologies” Programme of the Austrian Federal Ministry for Climate Action, Environment, Energy, Mobility, Innovation and Technology (BMK), the Austrian Federal Ministry of Labour and Economy (BMAW) and the Province of Styria. 
The COMET Programme is managed by the Austrian Research Promotion Agency (FFG).

\bibliography{SC_2023_literature}

\newcommand{\etalchar}[1]{$^{#1}$}
\begin{thebibliography}{BNGS{\etalchar{+}}20}

\bibitem[BKJ{\etalchar{+}}20]{BoTorch}
Maximilian Balandat, Brian Karrer, Daniel~R. Jiang, Samuel Daulton, Benjamin
  Letham, Andrew~Gordon Wilson, and Eytan Bakshy.
\newblock {BoTorch: A Framework for Efficient Monte-Carlo Bayesian
  Optimization}.
\newblock In {\em Advances in Neural Information Processing Systems 33}, 2020.

\bibitem[BNGS{\etalchar{+}}20]{Benajes2020}
J.~Benajes, R.~Novella, J.~Gomez-Soriano, I.~Barbery, C.~Libert,
  F.~Rampanarivo, and M.~Dabiri.
\newblock Computational assessment towards understanding the energy conversion
  and combustion process of lean mixtures in passive pre-chamber ignited
  engines.
\newblock {\em Applied Thermal Engineering}, 178:115501, 2020.

\bibitem[Fra18]{Frazier2018}
Peter~I. Frazier.
\newblock {A Tutorial on Bayesian Optimization}, 2018.
\newblock {\tt arXiv:1807.02811}.

\bibitem[FZQ{\etalchar{+}}18]{Feng2018}
Liyan Feng, Jun Zhai, Chuang Qu, Bo~Li, Jiangping Tian, Lei Chen, Weiyao Wang,
  Wuqiang Long, and Bin Tang.
\newblock The influence of the enrichment injection angle on the performance of
  a pre-chamber spark ignition natural-gas engine.
\newblock {\em Proceedings of the Institution of Mechanical Engineers, Part D:
  Journal of Automobile Engineering}, 232(5):679--694, 2018.

\bibitem[GBV{\etalchar{+}}19]{Guelpa2019}
Elisa Guelpa, Aldo Bischi, Vittorio Verda, Michael Chertkov, and Henrik Lund.
\newblock Towards future infrastructures for sustainable multi-energy systems:
  A review.
\newblock {\em Energy}, 184:2--21, 2019.
\newblock Shaping research in gas-, heat- and electric- energy infrastructures.

\bibitem[GBY{\etalchar{+}}21]{Ge2021}
Haiwen Ge, Ahmad~Hadi Bakir, Suren Yadav, Yunseon Kang, Siva Parameswaran, and
  Peng Zhao.
\newblock Cfd optimization of the pre-chamber geometry for a gasoline spark
  ignition engine.
\newblock {\em Frontiers in Mechanical Engineering}, 6:599752, 2021.

\bibitem[GG21]{Goesnitzer2021}
Clemens G{\"o}{\ss}nitzer and Shawn Givler.
\newblock A new method to determine the impact of individual field quantities
  on cycle-to-cycle variations in a spark-ignited gas engine.
\newblock {\em Energies}, 14(14):4136, 2021.

\bibitem[GKK{\etalchar{+}}19]{GhaderiMasouleh2019}
M.~{Ghaderi Masouleh}, K.~Keskinen, O.~Kaario, H.~Kahila, S.~Karimkashi, and
  V.~Vuorinen.
\newblock Modeling cycle-to-cycle variations in spark ignited combustion
  engines by scale-resolving simulations for different engine speeds.
\newblock {\em Applied Energy}, 250:801--820, 2019.

\bibitem[GKX{\etalchar{+}}14]{Gardner2014}
Jacob~R Gardner, Matt~J Kusner, Zhixiang~Eddie Xu, Kilian~Q Weinberger, and
  John~P Cunningham.
\newblock Bayesian optimization with inequality constraints.
\newblock In {\em Proc. Int. Conf. on Machine Learning (ICML)}, pages 937--945,
  2014.

\bibitem[GPP22]{Gossnitzer2022}
Clemens G{\"o}{\ss}nitzer, Stefan Posch, and Gerhard Pirker.
\newblock Development of a framework for internal combustion engine simulations
  in openfoam.
\newblock In {\em 8th European Congress on Computational Methods in Applied
  Sciences and Engineering: ECCOMAS CONGRESS 2022}, 2022.

\bibitem[NPGS{\etalchar{+}}20]{Novella2020}
Ricardo Novella, Jose Pastor, Josep Gomez-Soriano, Ibrahim Barbery, Cedric
  Libert, Fano Rampanarivo, Chistou Panagiotis, and Maziar Dabiri.
\newblock Experimental and numerical analysis of passive pre-chamber ignition
  with egr and air dilution for future generation passenger car engines.
\newblock In {\em WCX SAE World Congress Experience}. SAE International, apr
  2020.

\bibitem[Opea]{OpenFOAMCommit}
{OpenFOAM Development Repository}.
\newblock Github commit c95d964 from 20th November 2022,
  \url{https://github.com/OpenFOAM/OpenFOAM-dev}.
\newblock Accessed: 2023-01-03.

\bibitem[Opeb]{OpenFOAM}
{The OpenFOAM Foundation}.
\newblock \url{https://openfoam.org/}.
\newblock Accessed: 2023-01-03.

\bibitem[PGO{\etalchar{+}}22]{Posch2022}
Stefan Posch, Clemens G{\"o}{\ss}nitzer, Andreas~B Ofner, Gerhard Pirker, and
  Andreas Wimmer.
\newblock Modeling cycle-to-cycle variations of a spark-ignited gas engine
  using artificial flow fields generated by a variational autoencoder.
\newblock {\em Energies}, 15(7):2325, 2022.

\bibitem[PWZ{\etalchar{+}}21]{Posch2021}
S.~Posch, H.~Winter, J.~Zelenka, G.~Pirker, and A.~Wimmer.
\newblock Development of a tool for the preliminary design of large engine
  prechambers using machine learning approaches.
\newblock {\em Applied Thermal Engineering}, 191:116774, 2021.

\bibitem[RF03]{Roethlisberger2003}
R.P. Roethlisberger and D.~Favrat.
\newblock Investigation of the prechamber geometrical configuration of a
  natural gas spark ignition engine for cogeneration: part ii. experimentation.
\newblock {\em International Journal of Thermal Sciences}, 42(3):239--253,
  2003.

\bibitem[RW06]{RasmussenWilliams}
Carl~Edward Rasmussen and Christopher~K.I. Williams.
\newblock {\em {Gaussian Processes for Machine Learning}}.
\newblock The MIT Press, 2006.

\bibitem[SMB{\etalchar{+}}22]{Silva2022}
Mickael Silva, Balaji Mohan, Jihad Badra, Anqi Zhang, Ponnya Hlaing, Emre
  Cenker, Abdullah~S. AlRamadan, and Hong~G Im.
\newblock Doe-ml guided optimization of an active pre-chamber geometry using
  cfd.
\newblock {\em International Journal of Engine Research},
  0(0):14680874221135278, 2022.

\bibitem[SSH{\etalchar{+}}20]{Silva2020}
Mickael Silva, Sangeeth Sanal, Ponnya Hlaing, Bengt Johansson, Hong~G. Im, and
  Emre Cenker.
\newblock Effects of geometry on passive pre-chamber combustion
  characteristics.
\newblock In {\em WCX SAE World Congress Experience}. SAE International, apr
  2020.

\bibitem[WSP{\etalchar{+}}20]{Winter2018}
Hubert Winter, Eduard Schne{\ss}l, Gerhard Pirker, Jan Zelenka, and Andreas
  Wimmer.
\newblock {Application of CFD simulation to optimize combustion in prechamber
  gas engines with port injection}.
\newblock In {\em Proceedings of the 6th European Conference on Computational
  Mechanics: Solids, Structures and Coupled Problems, ECCM 2018 and 7th
  European Conference on Computational Fluid Dynamics, ECFD 2018}, pages
  894--905, 2020.

\bibitem[ZALY22]{Zhu2022}
Sipeng Zhu, Sam Akehurst, Andrew Lewis, and Hao Yuan.
\newblock A review of the pre-chamber ignition system applied on future
  low-carbon spark ignition engines.
\newblock {\em Renewable and Sustainable Energy Reviews}, 154:111872, 2022.

\bibitem[ZYE{\etalchar{+}}20]{Zhang2020}
Anqi Zhang, Xin Yu, Nayan Engineer, Yu~Zhang, and Yuanjiang Pei.
\newblock Numerical investigation of pre-chamber jet combustion in a light-duty
  gasoline engine.
\newblock In {\em Internal Combustion Engine Division Fall Technical
  Conference}, volume 84034, page V001T06A013. American Society of Mechanical
  Engineers, 2020.

\end{thebibliography}

\end{document}